\documentclass{aastex}
\usepackage{emulateapj5}

\slugcomment{Received 3 March 2000; Accepted 9 May 2000}

\shorttitle{GG Tau circumbinary ring}
\shortauthors{Silber et al.}

\received{4 August 1988}
\accepted{23 September 1988}

\begin{document}

\title{Near-infrared imaging polarimetry of the GG Tau circumbinary ring\altaffilmark{1}}

\author{Joel Silber,\altaffilmark{2} Tim Gledhill,\altaffilmark{2} Gaspard Duch\^{e}ne,\altaffilmark{3} and Fran\c{c}ois M\'{e}nard\altaffilmark{4}}

\altaffiltext{1}{Based on observations made with the NASA/ESA \textit{Hubble Space Telescope}, obtained at the Space Telescope Science Institute, which is operated by AURA, Inc., under NASA contract NAS5-26555.}
\altaffiltext{2}{Department of Physical Sciences, University of Hertfordshire, Hatfield, Hertfordshire, AL10 9AB, U.K.}
\altaffiltext{3}{Laboratoire d'Astrophysique, Observatoire de Grenoble, Universit\'{e} Joseph Fourier, B.P. 53, 38041 Grenoble Cedex 9, France}
\altaffiltext{4}{Canada-France-Hawaii Telescope Corporation, PO Box 1597, Kamuela, HI 96743}

\received{4 August 1988}
\accepted{23 September 1988}

\begin{abstract}
We present 1 micron \textit{Hubble Space Telescope\/}/NICMOS resolved
imaging polarimetry of the \hbox{GG Tau} circumbinary ring.  We find
that the ring displays east-west asymmetries in surface brightness as
well as several pronounced irregularities, but is smoother than
suggested by ground-based adaptive optics observations.  The data are
consistent with a $37\arcdeg$ system inclination and a projected
rotational axis at a position angle of $7\arcdeg$ east of north,
determined from millimeter imaging.  The ring is strongly polarized,
up to $\sim50$\%, which is indicative of Rayleigh-like scattering from
sub-micron dust grains.  Although the polarization pattern is broadly
centrosymmetric and clearly results from illumination of the ring by
the central stars, departures from true centrosymmetry and the
irregular flux suggest that binary illumination, scattering through
unresolved circum\-\textit{stellar\/} disks, and shading by these
disks, may all be factors influencing the observed morphology.  We
confirm a $\sim0\farcs25$ shift between the inner edges of the NIR and
millimeter images and find that the global morphology of the ring and
the polarimetry provide strong evidence for a geometrically thick
ring.  A simple Monte Carlo scattering simulation is presented which
reproduces these features and supports the thick ring hypothesis.  We
cannot confirm filamentary streaming from the binary to the ring, also
observed in the ground-based images, although it is possible that
there is material inside the dynamically cleared region which might
contribute to filamentary deconvolution artifacts.  Finally, we find a
faint 5th point source in the \hbox{GG Tau} field which, if it is
associated with the system, is almost certainly a brown dwarf.
\end{abstract}



\section{Introduction}
\label{sec:intro}

GG Tau is one of the most studied systems known to possess a tidally
truncated circumbinary disk.  It is a double binary, in which all four
components display evidence of infrared excesses \citep{white99}
indicating the presence of unresolved circum\-\textit{stellar\/}
disks.  However, the most striking object in the system is the large
disk of gas and dust surrounding the northern \hbox{($\sim0\farcs25$)}
binary.

Since the circumbinary structure was first detected
\citep{beckwith90,sg92,kawabe93}, many of its physical parameters have
been constrained, particularly from the millimeter interferometry of
\citet{dgs94} and \citet[hereafter GDS]{gds99}.  The disk is tidally
truncated by the dynamical action of the binary at a radius of
\hbox{$\sim180$AU} from the binary center of mass and extends out to
at least 800AU. From their most recent millimeter continuum and CO
line flux measurements, GDS propose that 90\% of the dust component
of the circumbinary disk lies in a well-defined \textit{ring\/} (inner
radius \hbox{$\sim180$AU;} outer radius \hbox{$\sim260$AU)} within the
extended (800AU) disk. Their observations are consistent with an
almost circular Keplerian disk, rotating about the binary's center of
mass with a projected rotational axis at a position angle (PA) of
$7\arcdeg$ east of north, and viewed at an inclination of $37\arcdeg$
to face-on. This implies that the northern part of the ring is
closest to us (the `front') and that the southern part of the ring
corresponds to the `back'.

The first near-infrared (NIR) observations of the ring were obtained
with ground-based adaptive optics (AO) by \citet[hereafter
RRNGJ]{roddier96} who published deconvolved $J$\/, $H$\/, and $K$\/
band images of the ring.  Their data show a very clumpy ring and
suggest a large degree of anisotropy in its illumination.  The images
also show radial filaments \textit{inside\/} the dynamically cleared
cavity, extending from the binary to ring.  At the $J$\/ and $K$\/
bands, the ring appears incomplete (or at least \textit{very\/} faint)
at the back.  GDS found that when they registered their 1.4mm and
RRNGJ's $J\/$ band image, there is a $\sim$0\farcs25 shift of the
ring's NIR inner edge towards the binary.  This could be explained if
the dust ring is \textit{geometrically thick\/}, extending to a height
of $\sim$120AU at its tidally truncated inner radius.

In this paper, we present the first resolved imaging polarimetry of
the \hbox{GG Tau} system and the first space-based images of the
circumbinary ring.  We refer to the northern binary as \hbox{`GG Tau'}
and follow GDS in assuming a distance of 140AU to the Taurus
star-forming region \citep{elias78}.

\section{Observations and Data Reduction}
\label{sec:obsdr}


Imaging linear polarimetry data of \hbox{GG Tau} were obtained on 1998
April 3 with the Near Infrared Camera and Multi Object Spectrometer
(NICMOS) on the \textit{Hubble Space Telescope (HST)} (Proposal 7827).
Camera 1 (NIC1) was used to image \hbox{GG Tau} through the three
polaroid filters, POL0S, POL120S, and POL240S, of central wavelength
\hbox{1.0459\micron} \hbox{(1\micron)} at a pixel scale of 0\farcs043.
The detector was operated in MULTIACCUM mode enabling multiple
non-destructive read-out of the array during integration. Our data
were re-reduced using the {\sc stsdas\/} package {\sc nicproto} to
remove the `pedestal' and `shading' image anomalies
\citep{dickinson99}.  We calculated the Stokes parameters and
associated errors from our data using an {\sc idl} implementation of
the algorithm developed by \citet{sa99} and the polaroid coefficients
of \citet{hines98}.


A PSF subtraction or deconvolution is essential to reveal the faint
circumbinary material surrounding \hbox{GG Tau}, since the PSF wings
from the binary are highly extended and possess diffraction spikes
which fill the entire frame. We have used our own observations of the
\hbox{DF Tau} \textit{binary\/} to subtract the \hbox{GG Tau} PSFs. DF
Tau is the closest binary in our sample and its $\sim$0\farcs1
separation is at the \textit{HST}/NICMOS resolution limit.  The
relative flux and absolute location of the \hbox{GG Tau} and \hbox{DF
Tau} components were measured with iterative PSF fitting in
\textsc{daophot} \citep{stetson87}. We then created a `double' image
for each \hbox{GG Tau} polaroid by shifting and adding the original
image to itself with the relative flux and separation of the \hbox{DF
Tau} components.  Appropriately scaled \hbox{DF Tau} PSFs were then
subtracted from these new images. We are confident that the total flux
subtracted is correct to within the photometric errors, and that the
PSF registration is accurate since the diffraction spikes are
subtracted to almost undetectable levels away from the PSF cores.

\begin{figure*}[htbp]
  \begin{center}
    \epsscale{2.0}
    \plotone{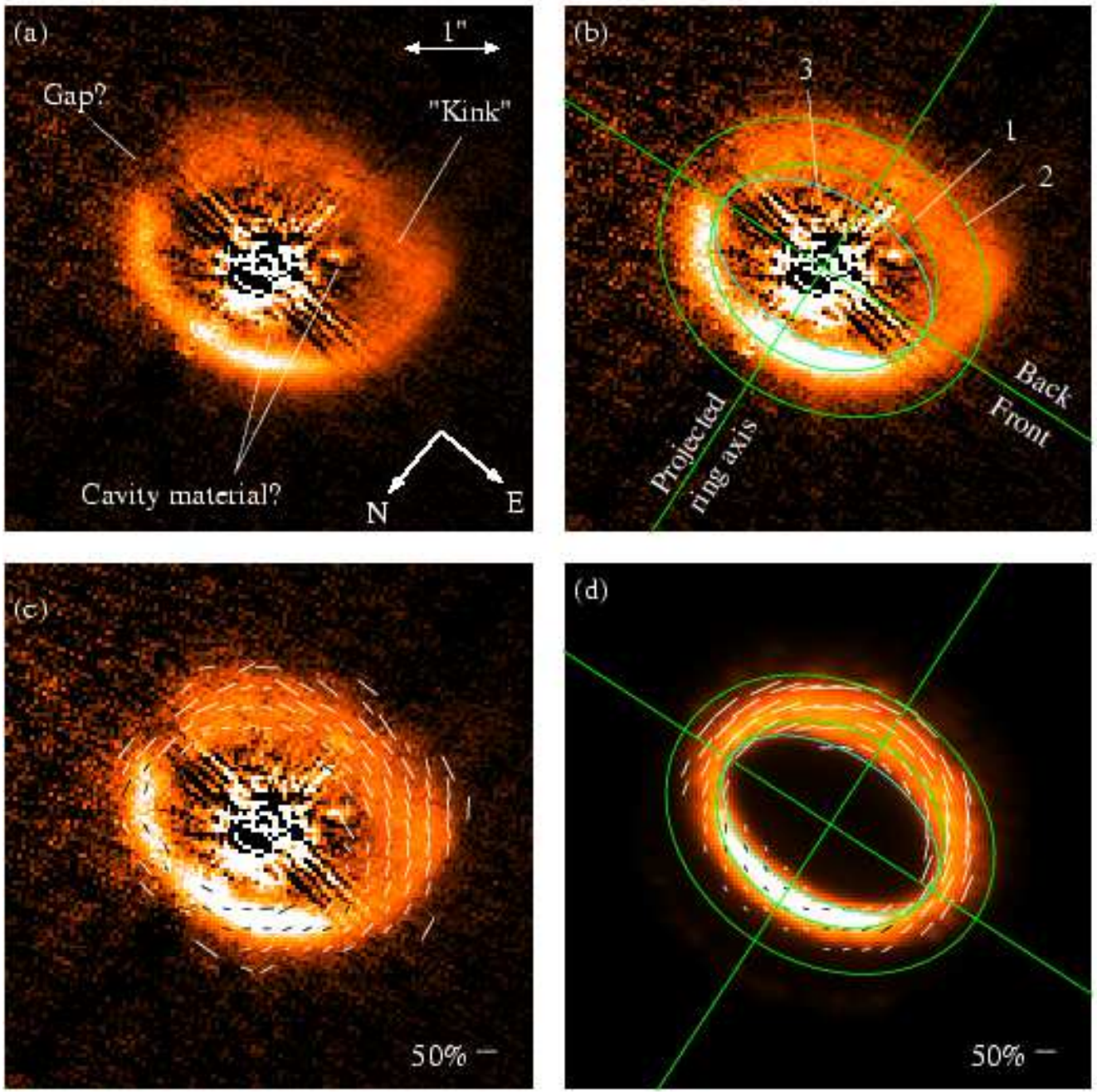}
    \caption{\textit{Left\/}: NIC1 PSF subtracted image of \hbox{GG Tau} at 1 micron revealing the circumbinary ring. Subtraction residuals, most obvious in the core and inner diffraction spikes remain due to the low sampling of the PSF image and errors in the relative separation and flux of the \hbox{GG Tau} and \hbox{DF Tau} binaries. The image is 5\farcs6 on a side $\approx$800AU assuming a distance of 140 parsec to Taurus (Elias 1978) and has been scaled linearly to reveal irregularities and asymmetries. (\textit{b\/}) and (\textit{c\/}): The same data overlaid with the projected rotational axis of the ring and ellipses based on the geometric findings of Guilloteau et al. (1999), and with linear polarization vectors. These images have been rescaled to highlight the northern bright arc. (\textit{d\/}): a simple Monte Carlo scattering simulation of the ring, overlaid with polarization vectors. The image has been convolved with the \hbox{DF Tau} binary to provide a like-for-like comparison with the data. \label{fig:ggd_color}}
  \end{center}
\end{figure*}

\section{Ring Morphology}
\label{sec:ringmorph}

A one micron PSF-subtracted NIC1 image of \hbox{GG Tau} is presented
in Figure \ref{fig:ggd_color}(a).  Although there are subtraction
residuals in the stellar cores and diffraction spikes, the ring is
exceptionally well revealed.  Unlike the smooth continuum flux
distribution at 1.4mm (GDS), and like RRNGJ's $J$\/ and $H$\/ band
images, the ring shows an unambiguous north-south asymmetry.  A
relatively thin bright arc of scattered light from the northern
(front) part of the ring dominates the image with broader and fainter
flux from the south (back) completing what is, essentially, an
unbroken ring.  The ring appears far less clumpy than in RRNGJ's
images.

The ring exhibits several examples of irregular structure: (i)
the arc is brighter on its east side; (ii) there is a `kink' in the
ring's south-east portion; (iii) there is a possible gap/dimming of
the ring (unfortunately) at the location of the western diffraction
spike.  Although GDS determined the large-scale distribution of dust
to be quite smooth, they used a $0\farcs9 \times 0\farcs6$ beam, and
it is possible that the irregularities we observe may be due to
small-scale inhomogeneities in the ring.  However, it is more likely
that we are witnessing the effects of anisotropic illumination of the
ring from the binary light sources and shading by the stars'
circumstellar disks.
There is some evidence of scattering from clumps of dusty material
\textit{inside\/} the dynamically cleared region at the locations of
\textit{both\/} of RRNGJ's $J$\/ and $H$\/ band filamentary streamers.
However, these clumps lie close to the subtraction residuals in our
image and may be residuals in their own right.  If real, it is
possible that RRNGJ's filaments are deconvolution artifacts caused by
actual material inside the ring.

\begin{figure*}
  \epsscale{0.7}
  \plotone{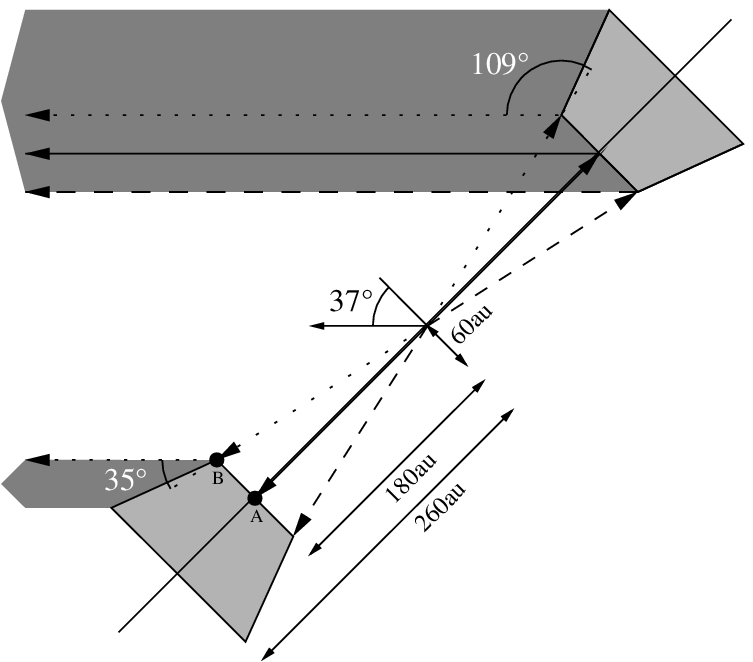}
  \caption{A representation of the thick ring geometry proposed by Guilloteau et al. (1999). Several light paths are drawn for single-scattering off the inner edges and surface of the \textit{optically\/} thick ring. The different projected surface areas from which scattered light is visible to an observer on the left result in very different thicknesses of observed flux around the ring (dark shading). This is exactly what is seen in our 1 micron image. This effect is reduced as one moves to thinner ring geometries (reduction of the disk height A$\rightarrow$B). In addition, this geometry predicts lower levels of polarization at the front of the ring than the back for Rayleigh-like scattering, since the scattering angles at the back ($109\arcdeg$) are closer to the maximum polarization scattering angle ($90\arcdeg$) than those at the front ($35\arcdeg$). This is also observed in the data.\label{fig:geometry}}
\end{figure*}

We have overlaid several ellipses on to our image in Figure
\ref{fig:ggd_color}(b) to help clarify the ring geometry.  These have
been located at the center of mass of the binary, assuming a 1:1 mass
ratio.  Ellipses 1 and 2 (green) represent circles of radii 180AU (1\farcs29)
and 260AU (1\farcs86), viewed at an inclination of $37\arcdeg$ to face
on, and should be interpreted as the projected millimeter (GDS) bounds
of the ring \textit{in its equatorial plane\/}.  Ellipse 3 (cyan) marks the 
projected ring/cavity boundary 
based on GDS's ring inclination, inner edge radius, and estimate of inner 
edge ring thickness (\S\ref{sec:intro}).
Scattered light is clearly observed inside ellipse 1, both at the
front and the back of the ring whereas we find that ellipse 3 provides 
a good bound
to the scattered flux edges, particularly at the front of the ring.  
Thus we are able to confirm the
$\sim$0\farcs25 displacement of the NIR front edge inwards towards the
binary with respect to the 180AU inner edge derived from 1.4mm
continuum imaging.  This is consistent with scattering occurring high
up in a geometrically thick ring, away from its equatorial plane.
Further evidence for this comes from the morphology of front and back
scattered flux.  Figure \ref{fig:geometry} demonstrates how an
optically and geometrically thick ring will produce the scattered flux
morphology of the observations -- the narrower flux at the front and
broader flux at the back.  In contrast, a geometrically thin ring
would produce similar widths of scattered flux from the front and back
of the ring, and an optically thin ring would produce homogeneous flux
similar to GDS's 1.4mm continuum image.  Knowledge of the ring's
orientation allows us to deduce that the thin bright arc must be
caused by forward-scattering off the ring's upper front surface and
the broad fainter flux results from back-scattering off the rear upper
surface and ring inner edge.

Following \citet{close98}, who re-examined the RRNGJ images, we
measured the ratio of total flux received from the front and back
parts of the ring.  Use of RRNGJ's `ring zone' (concentric ellipses
with semi-major axes 1\farcs25 and 1\farcs90) and rotational axis PA
($20\arcdeg$ east of north) does not make sense with our
image. Instead, the ratio was measured for all flux lying outside of
our ring inner bound, ellipse 3.  We obtain a value of $1.17\pm0.06$
which is considerably less than that measured by
\citet{close98} from RRNGJ's $J$\/ band image ($3.6\pm1.0$).  We have
also measured the azimuthal \textit{intensity\/} profile of flux
contained within two (non-concentric) ellipses, namely ellipse 3 and
an additional ellipse 0\farcs3 larger than ellipse 3 in both semi-axes
(not shown). These measurements reveal a maximum flux intensity ratio
of $\sim$4.0 between the peak of the brighter side of the arc and the
back of the ring. This contrasts with the $>$10:1 (2.8 magnitudes
difference) obtained by RRNGJ from their deconvolved $J$\/ band data.
This measurement provides an additional constraint for modelling the
system.

\section{Ring Polarization}
\label{sec:ringpol}

We present in Figure \ref{fig:ggd_color}(c) the first resolved linear
polarization map of the \hbox{GG Tau} ring.  At 1 micron the ring
displays a broadly centrosymmetric scattering pattern which clearly
indicates illumination of the ring by the central stars.  A detailed
interpretation of the polarization pattern will require inclusion of
binary sources and, probably, the effect of the binary's circumstellar
disks.  The high polarization levels observed at 1 micron, up to
$\sim50\%$, are indicative of Rayleigh-like scattering from sub-micron
particles.
The polarization is at a minimum ($\sim20\%$) in the bright arc at the
front of the ring and a maximum ($\sim50\%$) in the broad region at
the sides and back.  Using the geometry proposed by GDS, the
scattering angles involved are typically $\sim35\arcdeg$ at the front
and $\sim109\arcdeg$ (or greater) at the back (Figure
\ref{fig:geometry}).  Maximum polarization in the Rayleigh regime
occurs when photons are scattered through right-angles and,
consequently, since the back is scattering $36\arcdeg$ closer to
maximum polarization than the front, it is more highly polarized.
Thus our polarimetry observations are entirely consistent with a thick
disk geometry ($\S$1).  This behaviour would not be observed with a
thin disc where front and back scattering angles are equally close to
$90\arcdeg$.

\section{Simple Model}
\label{sec:model}

We performed some simple, single source, Monte Carlo multiple
scattering simulations to try and reproduce the ring's major
morphological and polarization features with a thick disk geometry.
We did not attempt any detailed calibration to previous determinations
of disk/ring mass, but sought only to demonstrate the effects of a
geometrically and optically thick ring on the scattering and
polarization patterns.  The ring was modelled as an arbitary standard
flared disk ($\alpha=2$ and $\beta=1.05$) with Gaussian truncated
inner and outer edges (parameters used by GDS).  The dust density
distribution parameters were all free, so there is bound to be much
degeneracy between absolute density, scale height, flaring parameter
$\beta$, and radial density power law $\alpha$.

Our best simulation is presented in Figure \ref{fig:ggd_color}(d) and
has been overlaid with its corresponding polarization vectors.
Following \citet{close98}, who implicitly modelled the ring as a flat
structure, we used the \citet{mw89} dust grain model `A' with a
steeper grain size power law ($p=-4.7$).  However, this simulation
retained the model's original maximum grain size of 0.9$\micron$.  The
major features of the observations are all reproduced with an integrated
front/back flux ratio of 1.0 and front/back maximum intensity ratio of 3.3
fitting the data well. The shift inwards of the inner edge of the ring
of $0\farcs25$, noted by GDS, is reproduced and is roughly consistent
with their ring thickness estimate at the inner edge of 120AU at a
radius of 180AU. The global polarization features (low at the front,
high elsewhere) are reproduced with polarization levels ranging from
10\% to 80\%, values broadly within our errors.  This is another
powerful endorsement of the geometrically thick disk hypothesis.

More thorough modelling of the ring will be the subject of a future
paper which will consider scattering through, and shading by, the
unresolved circumstellar disks, the effect of binary sources, and
choice of dust grain model, all of which will affect the scattering
and polarization patterns.  Finding the correct grain model is vital
if the ring is to be modelled successfully, as this determines both
the front/back flux ratios and the levels of polarization.  Further
observations of \hbox{GG Tau} are essential and, in particular, only 
multiple wavelength resolved polarimetry of the ring will provide 
enough constraints on the grain model to remove this as a free 
parameter.

\section{A Fifth Element?}
\label{sec:fifthelement}

We note the presence of a fifth point source in the \hbox{GG Tau}
field which is considerably fainter than the possibly substellar
\citep{white99} secondary of the \textit{southern\/} binary.  It is
located off the field of Figure \ref{fig:ggd_color}, 6\farcs16
($\sim$860AU) from the primary of the northern binary at a PA of
$242\arcdeg$ east of north.  Although the source appears in the
previously published coronagraphic images of \citet{ng95}, these
authors make no reference to it and we can find no comment regarding
the object in the literature.  We note that \hbox{GG Tau} does not lie
in the densest region of the Taurus-Auriga star-forming complex and
the object may well be a background star.  However, if it does lie at
the same distance as \hbox{GG Tau} then, given its faintness, it is
almost certainly a brown dwarf.

\acknowledgments

We wish to thank Dave Axon and Bill Sparks for early access to
preprints and their IDL polarimetry code, Eddie Bergeron for
assistance with image re-reduction, John Krist for advice on the PSF
subtractions, and Phil Lucas and Hiro Takami for assistance with the 
Monte Carlo simulations.  We acknowledge use of the data analysis facilities
provided by the Starlink Project which is run by CCLRC on behalf of
PPARC. An anonymous referee is thanked for useful comments.


\clearpage



\end{document}